\documentclass[aps,prl,reprint,superscriptaddress]{revtex4-2}

\usepackage{color}
\usepackage{mathtools}
\usepackage[normalem]{ulem}
\usepackage{graphicx}
\usepackage{caption}
\usepackage{subcaption}
\usepackage{dcolumn}
\usepackage{bm}
\usepackage{amsmath}
\usepackage{amssymb}
\usepackage{epsfig}
\usepackage{verbatim}
\usepackage{pstricks,pst-grad,pst-plot}
\usepackage{natbib}
\usepackage{url}
\usepackage{makeidx} 
\usepackage{float}\usepackage{epstopdf}
\usepackage{wasysym}
\usepackage{tikz}
\usepackage{tkz-graph}
\usepackage{pgfplots}
\usepackage[subnum]{cases}

\definecolor{crimson}{RGB}{220,20,60}
\definecolor{palevioletred1}{RGB}{225,130,171}
\definecolor{lightskyblue}{RGB}{135,206,250}
\definecolor{gold4}{RGB}{139,117,0}
\definecolor{lightgold1}{RGB}{238,220,130}

\begin{document}


\title{Multistability and High Codimension Bifurcations in  Synergistic Epidemics on Heterogeneous Networks}

\author{Francisco~J.~P{\'e}rez-Reche}
\affiliation{School of Natural and Computing Sciences, University of Aberdeen, Aberdeen, UK}
\email{fperez-reche@abdn.ac.uk}

\author{Sergei~N.~Taraskin}
\affiliation{St. Catharine's College and Department of Chemistry,
University of Cambridge, Cambridge, UK}
\email{snt1000@cam.ac.uk}

\begin{abstract}
We investigate the impact of network heterogeneity on synergistic contagion dynamics. By extending a synergistic contagion model to diverse heterogeneous network topologies, we uncover the emergence of novel dynamical regimes characterized by multiple stable states, separated by a rich set of bifurcations reaching up to codimension 4. Additionally, we demonstrate how synergy fundamentally reshapes the influence of nodes based on their degree. Unlike in non-synergistic epidemics, low-degree nodes can play a pivotal role in enabling network invasion at the onset of spread, while high-degree nodes can trigger explosive contagion. These findings challenge conventional control strategies, highlighting the need for new approaches to enhance or suppress synergistic contagion.
\end{abstract}


\maketitle


Network models have emerged as an indispensable framework for describing the dynamics of complex systems composed of interacting units \cite{Newman2010Networks:Introduction,Barrat2008DynamicalNetworks,Dorogovtsev2022TheNetworks}. These models have been widely applied to a variety of natural and artificial processes, including the spread of infections, information, products, and malware \cite{Pastor-Satorras2015,Castellano_RMP2009,Gleeson2013Binary-stateBeyond}. In particular, compartmental epidemic models on networks have provided valuable insight into epidemic invasion thresholds, the design of effective immunization strategies, and the impact of network structure on overall dynamics~\cite{Pastor-Satorras2015,Clusella2016b}.

Traditional compartmental epidemic models assume that the infection transmission rate, $\alpha$, between an infected-susceptible pair  is independent of the states of other neighboring nodes \cite{Pastor-Satorras2015}. Such models predict two possible infection regimes: (I) a disease-free regime when $\alpha < \alpha_c$, and (II) an epidemic regime when $\alpha > \alpha_c$. Although the critical transmission rate at the invasion threshold, $\alpha_c$, varies depending on the structure of the network, the transition between these two regimes remains continuous regardless of the topology of the network.

Pairwise interactions are often insufficient  to capture the complexity of real-world dynamics. Accurate modeling of processes such as infection transmission \cite{Silk2022CapturingSets}, biological invasions \cite{Ludlam2011ApplicationsSynergy,Gordon_Book2009}, and social contagion \cite{Centola2018,Hodas2014TheContagion} requires higher-order interactions involving groups of more than two nodes. Theoretical frameworks incorporating such interactions include threshold models \cite{Granovetter1978ThresholdBehavior,Watts_PNAS2002,chae_discontinuous_2015,Min2018CompetingContagion,Kook2021DoubleProcesses,Gleeson2007,Miller2016EquivalenceNetworks}, generalized epidemic models with memory \cite{Janssen_PRE2004,Dodds_PRL2004,Dodds2005AContagion.,Bizhani_PRE2012,Miller2016EquivalenceNetworks}, contagion on simplicial complexes \cite{Iacopini2019SimplicialContagion,Matamalas2020AbruptComplexes}, and synergistic epidemic models \cite{Perez_Reche_Synergy2011,Broder-Rodgers2015a,Taraskin2013a,GomezGardenes_PerezReche_SciRep2016,Liu_ChengLai_PRE2017,Taraskin2019BifurcationsGraphs,Assis-Copelli_PRE2009,Lin2023FromSpreading}. These models often reveal a third regime (III) at intermediate values of $\alpha$, where disease-free and epidemic states can coexist \cite{GomezGardenes_PerezReche_SciRep2016,Liu_ChengLai_PRE2017,Taraskin2019BifurcationsGraphs,Assis-Copelli_PRE2009,Lin2023FromSpreading,chae_discontinuous_2015,Min2018CompetingContagion,Kook2021DoubleProcesses,Gleeson2007,Miller2016EquivalenceNetworks,Iacopini2019SimplicialContagion,Matamalas2020AbruptComplexes}. In this regime, the likelihood of an epidemic is sensitive to the initial conditions of infection and explosive contagion can occur.

The observation of three distinct dynamical regimes (I-III) appears to be robust in the existing literature, emerging across various spreading processes, higher-order interactions, and network topologies. This raises the intriguing conjecture that three regimes may be a characteristic feature of synergistic spread on any network, just as two regimes (I and II) universally describe non-synergistic spread. 

Here, we show that synergistic epidemics on heterogeneous networks can exhibit more than three distinct dynamical regimes, highlighting the complex interplay between network topology and contagion. Identifying these regimes advances our understanding of complex epidemic behavior in real-world systems where synergy and heterogeneity are ubiquitous.

\paragraph{The model.} The model proposed in Ref.~\cite{Taraskin2019BifurcationsGraphs} for SIS synergistic epidemics spreading in random regular graphs is extended here to accommodate epidemics spreading in networks with arbitrary  node degree distribution. The nodes of the network can be either infected or susceptible. In a time interval $\delta t$, infected individuals can recover and become susceptible with probability $\mu \delta t$. Assuming that each of the $n$ infected neighbors of a susceptible node transmits the infection independently with the rate $\lambda_n$, the probability of infection can be expressed as $\Lambda_{n} \delta t=1-(1-\lambda_n \delta t)^n$. 

Synergistic transmission is captured through an explicit dependence of $\lambda_n$ on the number of infected neighbors of the recipient node. Following Refs.~\cite{GomezGardenes_PerezReche_SciRep2016,Taraskin2019BifurcationsGraphs}, we consider the S-synergy associated with synergistic effects between the susceptible neighbors of the recipient node. For instance, this concept is relevant to the spread of social content~\cite{GomezGardenes_PerezReche_SciRep2016}. Specifically, we assume $\lambda_{n} = \min \{\alpha e^{\beta(k-n)},1/\delta t\}$, where $k$ is the node degree of the recipient node. The parameter $\alpha$ represents the intrinsic transmission rate between an isolated infected-susceptible pair, while $\beta$ controls the degree of synergistic opposition ($\beta<0$) or reinforcement ($\beta>0$) of transmission by susceptible neighbors. 
Synergistic opposition is common in social settings: individuals may adopt protective behaviors, like mask-wearing, when surrounded by others who do. Similarly, the adoption of a product or idea may be discouraged if few peers have embraced it. Synergistic reinforcement can arise when the absence of infections or adopters signals safety or opportunity, encouraging risk-taking or early adoption.

To study the impact of network heterogeneity on synergistic epidemics, we analyse model and real networks. As model networks, we considered three distinct types of random graphs that span a range of degree heterogeneity: Erdos-Renyi (ER), binary (BG), and power-law (PL) networks, each characterized by different node degree distributions:
\begin{equation}
 \rho_k=\begin{cases}
        \frac{\langle k\rangle^k e^{-\langle k\rangle}}{k!}& :\textrm{ ER$(\langle k\rangle)$},\\
         \varphi \delta_{kk_1}+(1-\varphi)\delta_{kk_2} & :\textrm{ BG$(k_1,k_2, \varphi)$},\\
         \frac{k^{-m}}{A}, k\in [k_{\min},  k_{\max}] & :\textrm{ PL$(k_{\min},k_{\max},m)$}.
        \end{cases}
        \label{eq:ProbDist_k}
\end{equation}
Here, $\delta_{kk'}$ denotes the Kronecker delta, $A=\sum_{k=k_{\min}}^{k_{\max}} k^{-m}$ is a normalization factor, and $\langle k \rangle$ is the mean node degree. For real networks, we analysed four datasets available from \cite{Kunegis2013KONECTCollection}: the Douban social network, the Western US power grid \cite{Rossi2015TheVisualization}, a human proteins network \cite{Rual2005TowardsNetwork} and the co-purchase network of Amazon \cite{Yang2015DefiningGround-truthb}.

Our model can describe both continuous ($\delta t \rightarrow 0$) and discrete-time dynamics (finite $\delta t$) with the results mainly presented for the latter which encompasses all the effects observed for continuous dynamics (see Figs. 2(a) and 4(a) in \cite{SM} for continuous-time examples). Specifically, we set $\delta t=1$ so that $\alpha \in[0,1]$ becomes an intrinsic transmission probability. Within this framework, the spreading dynamics depends on a set of parameters $\Gamma = \Gamma_E \cup \Gamma_N$, where $\Gamma_E = \{ \alpha,\beta,\mu \}$ are epidemic-related parameters and $\Gamma_N$ are parameters describing the node degree distribution
(e.g. $\Gamma_N = \{k_1,k_2,\varphi\}$ for a binary graph).

We studied the model in the long-time limit ($t \rightarrow \infty$), which determines whether the network is vulnerable to spread of infection or not. To gain mathematical insight, a single-site heterogeneous mean-field (SSHMF) approximation was employed. The SSHMF ignores dynamical correlations between the states of connected nodes and assumes that the probability that a node is infected only depends on its degree $k$~\cite{Gleeson2013Binary-stateBeyond,Pastor-Satorras2015}. The results obtained by this approach are qualitatively consistent with Monte Carlo simulations of the exact model, which we run on networks constructed with the uncorrelated configuration model~\cite{Catanzaro2005Generation027103,Newman2010Networks:Introduction} (see results in \cite{SM}).

\paragraph{SSHMF approximation.} Within the SSHMF, the probability $p_k(t;\Gamma)$ that a node with $k$ neighbours is infected is the solution of the following equation:
\begin{equation}
p_k(t+1;\Gamma)= (1-\mu) p_k(t;\Gamma)+(1-p_k(t;\Gamma))q_k(\theta)~,
\label{eq:pkt}
 \end{equation}
where 
\begin{equation}
 q_k(\theta)=\sum_{n=1}^{k} {k \choose n} {\Lambda}_n \theta^n (1-\theta)^{k-n}~,
\end{equation}
is the rate at which a susceptible node with $k$ neighbors becomes infected. This rate depends on the probability 
\begin{equation}
\theta = \frac{\langle k p_k \rangle}{\langle k \rangle}\,,
\label{eq:theta_general}
\end{equation}
that any given neighbor of the recipient node is infected. Here, $\langle \cdot \rangle$ denotes the average over the node degree distribution.

The predictions of the model for $p_k$ in the long-time limit correspond to the fixed points of Eq.~\eqref{eq:pkt} which gives
\begin{equation}
p_k^{\infty}(\Gamma) \equiv \lim_{t \rightarrow \infty} p_k(t;\Gamma) = f_k(\theta,\Gamma) \equiv \frac{q_k(\theta)}{\mu+q_k(\theta)}\,,
 \label{pk_equil}
\end{equation}
with $\theta$ being the zeros of the following function (cf. Eq.~\eqref{eq:theta_general}):
\begin{equation}
 F(\theta;\Gamma)=-\theta+\frac{\langle k f_k(\theta) \rangle}{\langle k \rangle}\,.
 \label{Ftheta}
\end{equation}

\begin{figure}
\includegraphics[clip=true,width=8cm]{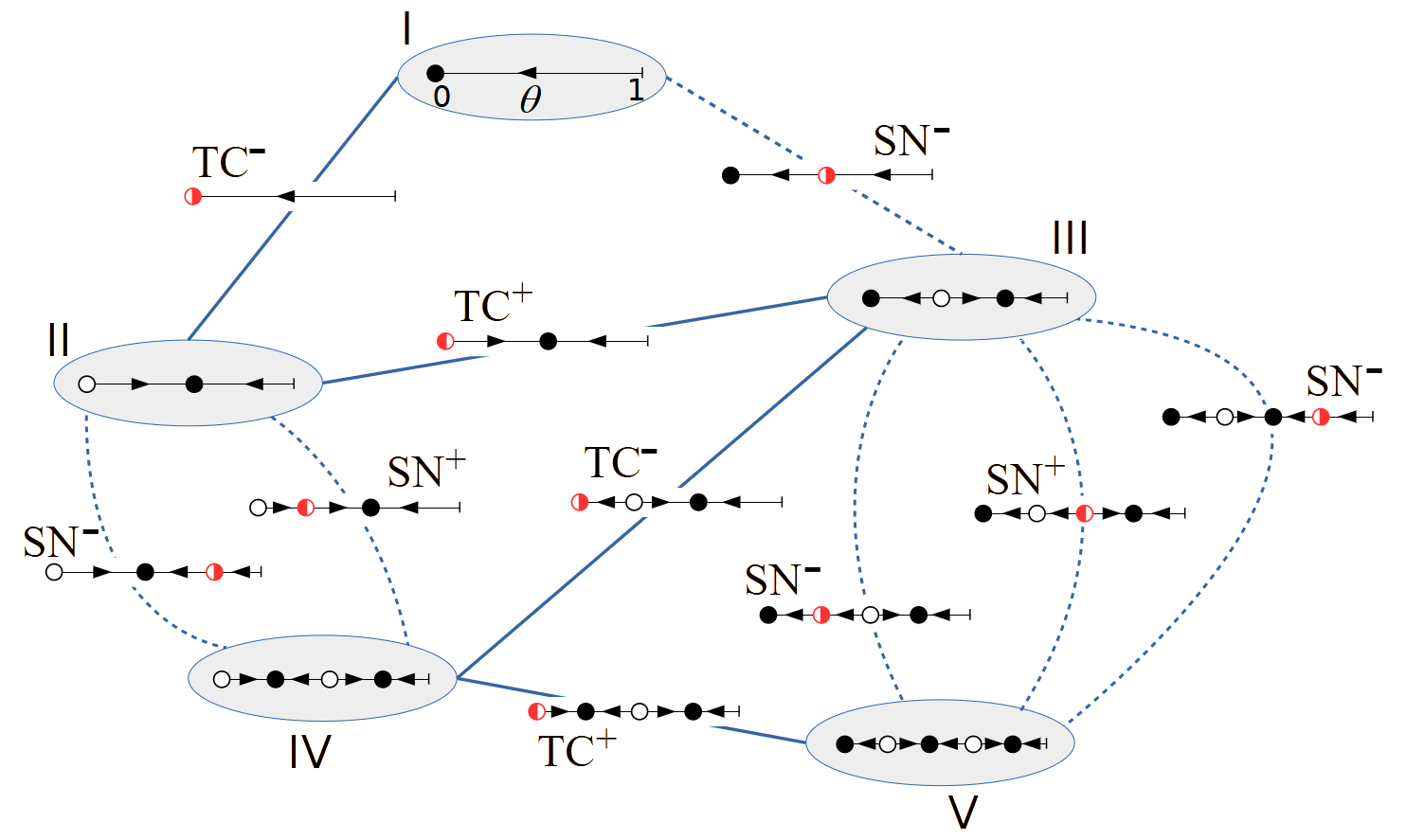}
 \caption{\label{fig:Bifurcation_graph} Graph representation of the five observed dynamical regimes (nodes represented by ellipses) and the codimension-1 bifurcations connecting them (solid and dashed links for TC$^{\pm}=\mathcal{B}_{1}^{\pm}(0)$ and SN$^{\pm}=\mathcal{B}_{1}^{\pm}(\theta_*)$, respectively). Flow diagrams illustrate the dynamics of $\theta$, with arrows indicating evolution toward stable ($\CIRCLE$) or away from unstable ($\Circle$) fixed points.  Bifurcation points are marked by half-filled circles: $\RIGHTcircle$ for negative and $\LEFTcircle$ for positive bifurcations.
 } 
\end{figure}

A fixed point at $\theta_*$ is stable if $\partial_{\theta} F(\theta;\Gamma)|_{\theta=\theta_*}<0$. The infection-free state with $\theta_*=0$ is always a fixed point since $F(0;\Gamma)=0$ for any $\Gamma$. Additionally, for the networks considered here, $F(\theta;\Gamma)$ can have up to four positive zeros. Consequently, the long-time behavior of the model can fall into five different regimes (Fig.~\ref{fig:Bifurcation_graph}) distinguished by the total number of fixed points. Regime I represents an infection-free stable state characterized by $\theta_*=0$. 
In regimes II-V, there is at least one stable fixed point located at $\theta_*>0$, corresponding to endemic infection. In regimes III and V, both the infection-free and endemic states are stable. In regimes IV and V, 
two stable endemic states are possible depending on the initial conditions.

Variation in the model parameters can induce transitions between regimes, which are associated with bifurcations where the number of fixed points changes. A codimension-$\bar{d}$ bifurcation occurs at $\theta = \theta_* \in [0,1)$ if a non-empty set, $\mathcal{B}_{\bar{d}}^s(\theta_*)$, of parameters exists, defined as follows:
\begin{widetext}
\begin{equation}
 \mathcal{B}_{\bar{d}}^s(\theta_*) = \left\{\Gamma \middle| F(\theta_*;\Gamma) =0, \{ F^{(l)}(\theta_*;\Gamma) =0 \}_{l=1}^{\bar{d}}, F^{(\bar{d}+1)}(\theta_*;\Gamma) \neq 0 \right\}~.
\end{equation}
\end{widetext}
Here, $F^{(l)}(\theta_*;\Gamma)$ denotes the $l$-th derivative of $F$ with respect to $\theta$, evaluated at $\theta_*$, and $s=\text{sign}(F^{(\bar{d}+1)}(\theta_*;\Gamma))$.  The codimension $\bar{d}$ is the minimum number of parameters that must be varied for the bifurcation to occur \cite{Strogatz1994NonlinearChaos}.
Following this, the maximal codimension of bifurcations in a system with parameter set $\Gamma$ is $\bar{d}_{\max} = |\Gamma|$, where $|\cdot|$ indicates the number of parameters.

For the heterogeneous networks studied here, we observed bifurcations up to codimension-4 that connect the long-time regimes I-V. Specifically, Fig.~\ref{fig:Bifurcation_graph} illustrates transitions associated with codimension-1 bifurcations. Non-synergistic epidemics ($\beta=0$) can only exhibit regimes I and II irrespective of the node degree distribution (see Sec. I in \cite{SM}). Regimes I and II are connected by a transcritical bifurcation, TC$^- = \mathcal{B}_{1}^{-}(0)$, which drives the well-known continuous transition between infection-free and endemic states. 

For synergistic epidemics on regular graphs, a discontinuous transition between regimes I and III can occur, driven by a saddle-node bifurcation, SN$^{-} \equiv \mathcal{B}_{1}^{-}(\theta_*)$, located at $\theta^*\in (0,1)$ \cite{Taraskin2019BifurcationsGraphs}. The TC$^-$ and SN$^{-}$ bifurcations collide at a codimension-2 saddle-node-transcritical bifurcation, SNT$^- \equiv \mathcal{B}_{2}^{-}(0)$. 

\paragraph{Codimension-1 and 2 bifurcations in heterogeneous networks.}

\begin{figure*}
{\includegraphics[clip=true,width=12.75cm]{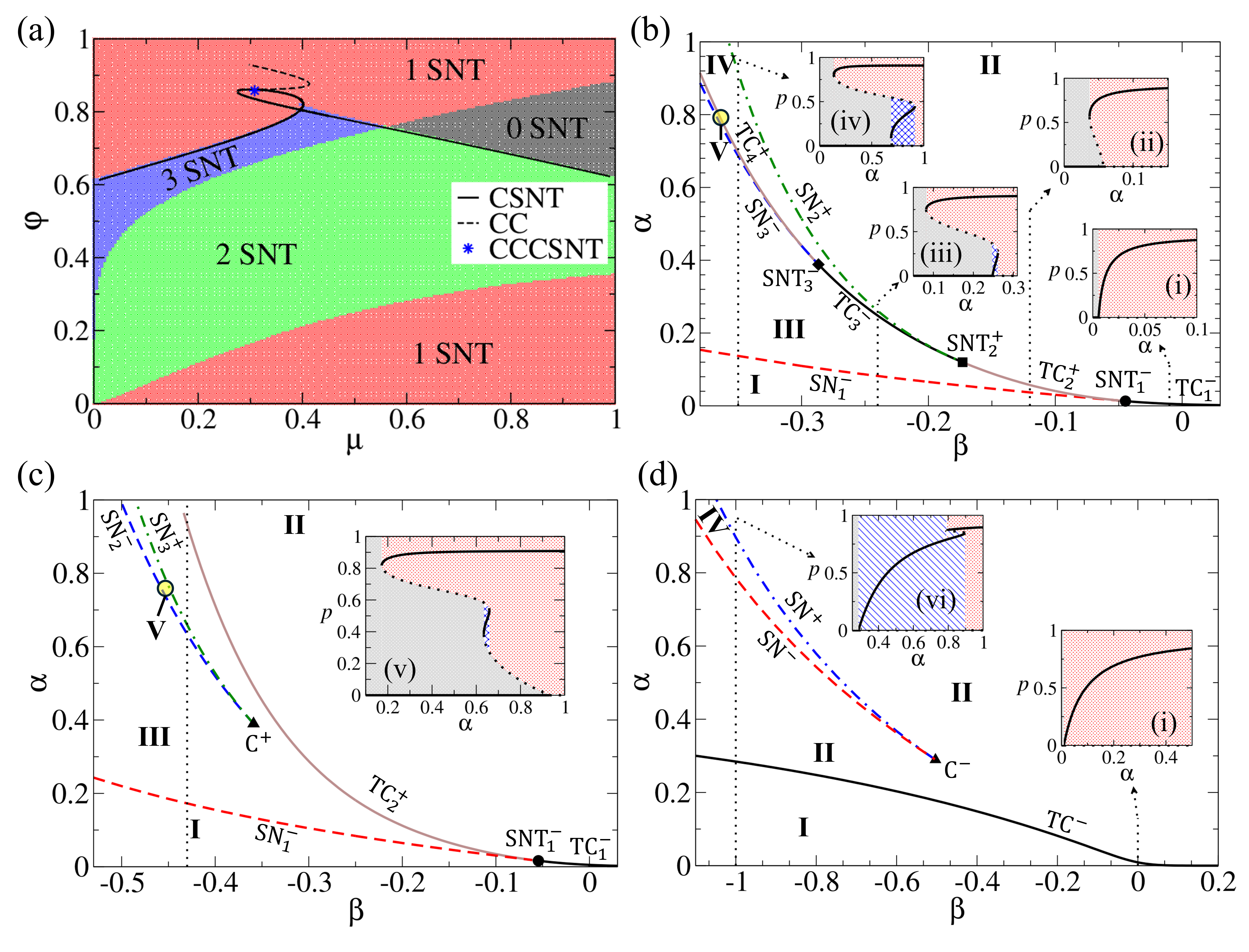}}

 \caption{
   Bifurcation diagrams and the mean proportion $p$ of infected nodes in synergistic SIS epidemics. (a) Codimension-4 bifurcation (CCCSNT$^-$, star) in BG$(10,30,\varphi)$, projected onto the $(\mu,\varphi)$ plane at $(\alpha,\beta,\mu,\varphi)=(0.096,-0.104,0.298,0.861)$. Codimension-3 bifurcations (CSNT, CC) are shown as solid and dashed lines, respectively. Colors indicate regions with different numbers of codimension-2 SNT bifurcations.
(b,c) Bifurcations in the $(\alpha,\beta)$ plane for $\mu=0.1$ on BG$(10,30,\varphi)$ with $\varphi=0.6$ and $0.76$, respectively. Regimes I–V are identified and invasion curves $p(\alpha)$ for regime transitions (i)–(v) are shown in the insets. Stable/unstable states shown with solid/dashed lines. Shaded regions represent basins of attraction: grey (infection-free), blue dashed (low-endemic), red dotted (high-endemic).
(d) Analogous results for $\mu=0.1$ on PL$(2,50,2)$ graphs; insets show transitions (i) and (vi).  In (b)-(d), codimension-1 bifurcations of TC and SN type are indicated by continuous and dashed lines, respectively.
}
   \label{fig:Bifurcations_alpha_beta} 
\end{figure*}

In contrast to regular graphs, the synergistic spread in heterogeneous networks can exhibit multiple SNT bifurcations.  This phenomenon, which can be observed for all the network types studied here, is illustrated by the bifurcation diagram in $(\mu,\varphi)$ parameter space for binary graphs BG$(10,30,\varphi)$  shown in Fig.~\ref{fig:Bifurcations_alpha_beta}(a) (see examples for ER, PL and real networks in \cite{SM}). 

Fig.~\ref{fig:Bifurcations_alpha_beta}(b) shows a bifurcation diagram in the $(\alpha,\beta)$ space for epidemics with $\mu=0.1$ spreading on a $BG(10,30,0.6)$ which exhibit three SNT bifurcations. Depending on the values of $\alpha$ and $\beta$, the long-time behavior of epidemics can fall into any of the five regimes I to V. As expected, for any $\beta$ and sufficiently small $\alpha$ the epidemic does not spread. The specific regimes observed for intermediate values of $\alpha$ depend on the value of $\beta$.  The insets in Fig.~\ref{fig:Bifurcations_alpha_beta}(b) show the expected proportion of infected nodes, $p=\langle p_k^{\infty} \rangle$, as a function of $\alpha$ for four different values of $\beta$. These invasion curves illustrate different ways to transition between infection-free and endemic regimes through codimension-1 bifurcations, corresponding to different paths in the graph of Fig.~\ref{fig:Bifurcation_graph}: 

\begin{description}
\item[(i)] $\beta>\beta_{\text{SNT}_1^-}$: A classical smooth transition between regimes I and II occurs at a TC$_1^-$ bifurcation as $\alpha$ increases.

\item[(ii)] $\beta \in (\beta_{\text{SNT}_2^+},\beta_{\text{SNT}_1^-})$: Three regimes, I-III, are possible. 
Increasing $\alpha$ from regime I leads to a discontinuous transition to regime II at TC$_2^+$. Conversely, decreasing $\alpha$ from regime II results in a discontinuous transition back to regime I at SN$_1^-$. In regime III, the system can settle into either the infection-free or endemic state depending on initial conditions, as illustrated by the basins of attraction in Fig.~\ref{fig:Bifurcations_alpha_beta}(b).

\item[(iii)] $\beta \in (\beta_{\text{SNT}_3^-},\beta_{\text{SNT}_2^+})$: 
Regimes I-IV can be observed. As $\alpha$ increases from regime I, $p$ initially rises smoothly at TC$_3^-$, followed by a discontinuous jump to regime II at SN$_2^+$. Decreasing $\alpha$ from regime II leads to a discontinuous transition back to regime I at SN$_1^-$.
\item[(iv)] $\beta < \beta_{\text{SNT}_3^-}$: 
The regime transition sequence I-III-V-IV-II is observed as $\alpha$ increases. These regimes are separated by the sequence of bifurcations, SN$_1^{-}$, SN$_3^{-}$, TC$_4^+$, and SN$_2^{+}$.
Increasing $\alpha$ from regime I leads to two discontinuities in $p$, separated by an interval of smooth variation corresponding to endemic states with infection levels lower than those in regime II. 
As in case (iii), decreasing $\alpha$ results in a discontinuous transition back to regime I at SN$_1^-$.
\end{description}

In addition to SNT bifurcations, synergistic epidemics in heterogeneous networks can exhibit codimension-2 cusp bifurcations, denoted as C$^{\pm} \equiv \mathcal{B}_{2}^{\pm}(\theta_*)$, which occur for $\theta_* \in (0,1)$ and result from the SN$^+$-SN$^-$ interaction. Fig.~\ref{fig:Bifurcations_alpha_beta}(c) presents a bifurcation diagram for a binary graph $BG(10,30,0.82)$, displaying two codimension-2 bifurcations: SNT$_1^-$ and cusp C$^+$. When $\beta>\beta_{\text{C}^+}$, varying $\alpha$ at fixed $\beta$ leads to transitions between regimes I and II through the codimension-1 bifurcations previously described, following regime sequences of type (i) or (ii). In contrast, 
\begin{description}
\item[(v)] if $\beta<\beta_{\text{C}^+}$, the system can fall sequentially into regimes  I-III-V-III-II separated by the sequence of bifurcations, SN$^-$, SN$^-$, SN$^+$ and TC$^+$ (inset in Fig.~\ref{fig:Bifurcations_alpha_beta}(c)). In this case, all transitions between different regimes are discontinuous. 
\end{description}

A codimension-2 bifurcation of type C$^{-}$ can also exist without SNT bifurcations, thus yielding  a qualitatively distinct bifurcation diagram, as illustrated in Fig.~\ref{fig:Bifurcations_alpha_beta}(d) for PL networks; similar results for real networks are shown in \cite{SM}.
When $\beta>\beta_{\text{C}^{-}}$, regimes I and II are connected by a smooth transition associated with a TC$^-$ bifurcation, resembling  case (i) described above. In contrast,
\begin{description}
\item[(vi)] if $\beta<\beta_{\text{C}^{-}}$, the regime sequence I-II-IV-II can be observed (see inset in Fig.~\ref{fig:Bifurcations_alpha_beta}(d)). Increasing $\alpha$ from regime I at constant $\beta$ leads to a gradual increase in $p$ at TC$^-$, followed by a discontinuous transition to regime II at SN$^+$. In contrast, decreasing $\alpha$ from regime II results in a discontinuous drop in $p$ at SN$^-$, followed by a smooth decrease toward regime I. 
\end{description}

\paragraph{Codimension-3 and 4 bifurcations in heterogeneous networks.}

Codimension-3 bifurcations of type $\mathcal{B}_{3}^{\pm}(0)$ can occur for epidemics spreading on binary, ER, PL graphs and real networks. These bifurcations arise from the interaction of C and SNT bifurcations, and are referred to as cusp-(saddle-node-transcritical) (CSNT) bifurcations. The solid curve in Fig.~\ref{fig:Bifurcations_alpha_beta}(a) shows the locus of such bifurcations in the space $(\mu,\varphi)$ for epidemics spreading on $BG(10,30,\varphi)$ graphs.

When crossing a CSNT$^{-}$ bifurcation by increasing $\alpha$, the proportion of infected nodes, $p$, increases smoothly, resembling the behavior observed when crossing $TC_1^-$ in case (i) described above (see Fig.~1(e) in \cite{SM}). 
Crossing a CSNT$^{+}$ bifurcation by increasing $\alpha$ results in a discontinuous jump in $p$, analogous to the behavior seen when crossing $TC_2^+$ in case (ii) (see Fig.~1(d) in \cite{SM}).

The only bifurcation of codimension-3 observed at $\theta_*\in (0,1)$ is of type $\mathcal{B}_{3}^{-}(\theta_*)$, and it occurs for epidemics spreading on binary graphs. These bifurcations, referred to as cusp-cusp (CC$^-$) bifurcations, result from the merging of two cusp bifurcations. For $BG(10,30,\varphi)$, they are located along the dashed line shown in Fig.~\ref{fig:Bifurcations_alpha_beta}(a). When crossing a CC$^-$ bifurcation by decreasing $\alpha$, $p$ exhibits a discontinuous drop from regime III to regime I, analogous to the behavior observed when crossing a SN$^-$ bifurcation in cases (ii)-(v) (see Figs.~1(b)-(c) in \cite{SM}).

A single codimension-4 bifurcation was identified in the networks studied: a $\mathcal{B}_{4}^{-}(0)$ bifurcation termed a (cusp-cusp)-(cusp-saddle-node-transcritical) (CCCSNT$^-$) bifurcation. This bifurcation results from the interaction between CSNT$^+$ and CC$^-$ bifurcations. The CCCSNT$^-$ bifurcation occurs in epidemics spreading on binary graphs with only a specific choice of parameters $(\alpha, \beta, \mu, \varphi)$ (see the star symbol in Fig.~\ref{fig:Bifurcations_alpha_beta}(a)). Crossing the CCCSNT$^-$ bifurcation by increasing $\alpha$ leads to a smooth transition between regimes I and II (see Fig.~1(a) in \cite{SM}). No codimension-4 bifurcations were observed for epidemics spreading in other studied networks, as the number of free parameters was less than four, implying $\bar{d}_{\max} < 4$. 

\paragraph{Additional regimes and bifurcations.}
Additional regimes and bifurcations beyond codimension-4 may arise for specific synergy parameters and degree distributions. This can be explored by extending our analysis to configuration-model networks with fixed degree support $\mathcal{K} \subset \mathbb{Z}^+$ and $\rho_k=\sum_{k'\in \mathcal{K}} \varphi_{k'} \delta_{kk'}$, characterized by $|\Gamma_N| = |\mathcal{K}| - 1$ free parameters (due to normalization, $\sum_{k\in \mathcal{K}} \varphi_k=1$). The node degree distribution of any real network can be seen as a particular realization of this model; binary graphs correspond to the case $\mathcal{K} = \{k_1, k_2\}$. For discrete-time synergistic epidemics with $\Gamma_E = \{\alpha, \beta, \mu\}$, bifurcations up to codimension $\bar{d}_{\max} = |\Gamma_E| + |\Gamma_N| = 2 + |\mathcal{K}|$ may occur.

The number of dynamical regimes depends on the number of zeros of $F(\theta;\Gamma)$ in Eq.~\eqref{Ftheta}, which must be evaluated separately for each network.  
As shown in Sec. IV of \cite{SM}, these regimes and bifurcations are highly sensitive to specific features of the degree distribution, precluding general predictions without detailed analysis.

\begin{figure}
{\includegraphics[clip=true,width=8.5cm]{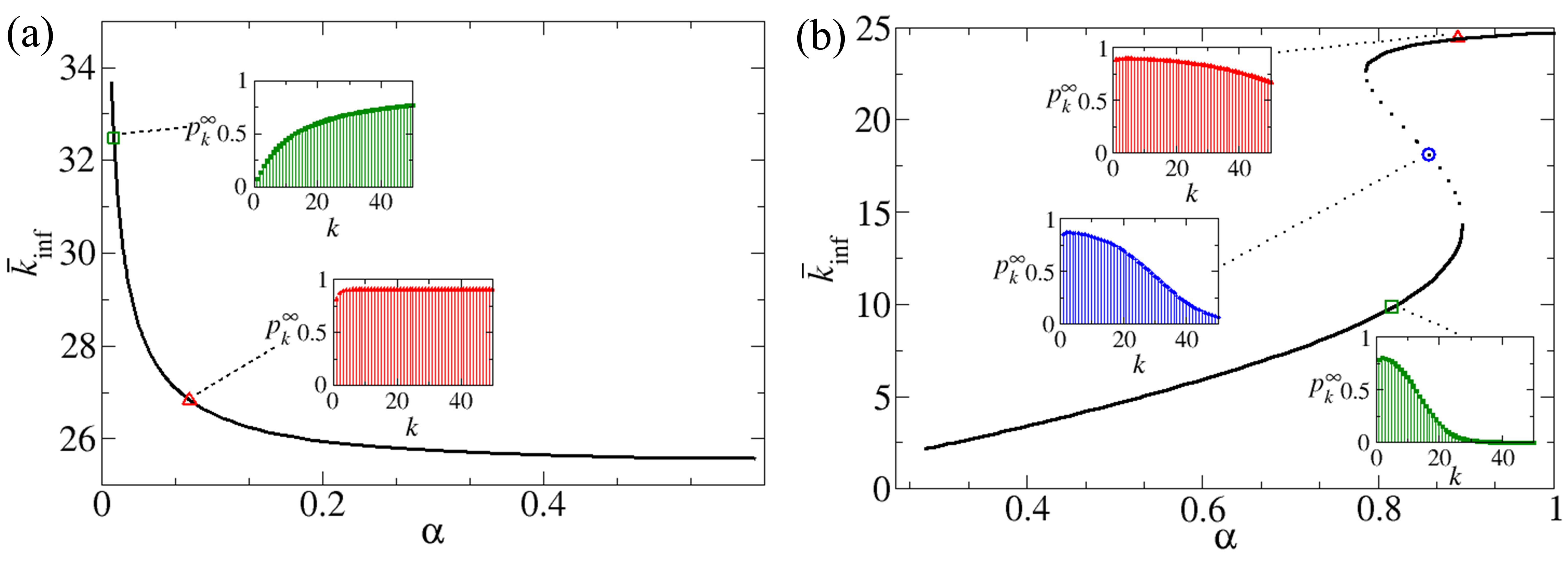}}
\caption{
  Mean degree of infected nodes $\bar{k}_{\text{inf}}$ vs. $\alpha$ in SIS epidemics with $\mu=0.1$ on PL$(1,50,2)$ graphs.
(a) Non-synergistic case ($\beta=0$), corresponding to invasion curve (i) in Fig.~\ref{fig:Bifurcations_alpha_beta}.
(b) Synergistic case ($\beta=-1$), corresponding to curve (iv) in the same figure.
Insets show the infection probability distribution $p_k^{\infty}$ at selected $\alpha$ values.
}
   \label{fig:kinf}
\end{figure}

\paragraph{Heterogeneous infection.}
For non-synergistic epidemics on heterogeneous networks, it is well established that the transition between disease-free regime I and endemic regime II is primarily driven by the infection of high-degree nodes \cite{Pastor-Satorras2015}.
Fig.~\ref{fig:kinf}(a) illustrates this for non-synergistic epidemics spreading on a PL network. When increasing $\alpha$ and crossing the TC$^-$ bifurcation, the system smoothly enters regime II. The average degree of infected nodes, $\bar{k}_{\text{inf}}=\sum_k k p_k^{\infty}/\sum_k p_k^{\infty}$,  decreases monotonically as nodes with a lower degree become more likely to be infected (see $p_k^{\infty}$ in the insets).  A similar pattern holds for synergistic epidemics with $\beta>0$. 

In contrast, negative synergy ($\beta<0$) leads to qualitatively different dynamics, shown in Fig.~\ref{fig:kinf}(b). Here, lower-degree nodes are more easily infected, and $\bar{k}_{\text{inf}}$ increases with $\alpha$. More notably, abrupt transitions in infection levels occur as high-degree nodes suddenly switch between susceptible and infected states, corresponding to the SN$^{\pm}$ bifurcations. The insets show $p_k^{\infty}$ distributions for low- (green) and high- (red) endemic states, with unstable distributions in blue. Similar examples of $p_k^{\infty}$ distributions for PL networks are shown in Fig.~9 of \cite{SM}. 

\paragraph{Conclusions.} 
We have demonstrated that the combination of network heterogeneity and synergistic effects in epidemics can produce far more complex dynamics than those found in non-synergistic models or synergistic epidemics on homogeneous networks. Using a single-site mean-field approach, supported by exact numerical simulations, we identified five distinct long-term regimes linked by bifurcations of up to codimension four. Notably, the emergence of regimes IV and V allows for two stable low- and high-endemic states, a phenomenon previously unreported in this class of models.  Such regimes have been observed in threshold models, but only with heterogeneous susceptibility of the nodes \cite{Dodds2005AContagion.}. Recognizing multiple endemic states is crucial for designing context-dependent control strategies that could promote transitions to desirable levels of endemicity.

Our results reveal the contrasting roles of high- and low-degree nodes in epidemics with negative synergy ($\beta < 0$), where susceptible neighbors inhibit transmission. In contrast to non-synergistic spread---where high-degree nodes drive endemicity---low-degree nodes here sustain infections as $\alpha$ increases, while high-degree nodes are responsible for abrupt changes in prevalence. This underscores the counterintuitive effects of synergy on epidemic thresholds and informs the design of adaptive control strategies.

The model studied here is a special case of a broader class of models for spreading processes for which the mean proportion of infected nodes $p$ changes over time with the rate $R(p)$ obeying the following properties: (i) $R(0)=0$, ensuring $p$ remains unchanged in the absence of infection, and (ii) $R(1)<1$, guaranteeing that a fully infected system is unstable due to recovery. Nonlinearities in $R$ can give rise to multiple quasi-stationary regimes with $R=0$ and complex bifurcation diagrams. In synergistic epidemics on networks, we have shown that these emerge from the interplay between network heterogeneity and transmission dynamics. More broadly, we expect such phenomena to be widespread in systems with high-order interactions between nodes leading to nonlinear $R(p)$. 
In particular, analogous phenomena may appear in models where synergy stems from infected neighbors \cite{Taraskin2019BifurcationsGraphs}. Based on previous studies on synergistic epidemics with node removal on regular graphs \cite{GomezGardenes_PerezReche_SciRep2016}, we conjecture that similar bifurcation complexity could result from nonlinearities in the equation for the size of the population removed at the end of the epidemic.

\begin{acknowledgments} 
F.J.P.R. acknowledges funding from BBSRC (grant no. BB/Y513027/1). 
\end{acknowledgments}


%

\end{document}